\documentclass[twocolumn]{aastex63}

\usepackage{amsmath}
\usepackage{graphicx}
\usepackage{epstopdf}
\usepackage{enumerate}
\usepackage{soul}

\usepackage{savesym}
\savesymbol{tablenum}
\usepackage{siunitx}
\restoresymbol{SIX}{tablenum}

\usepackage{hyperref}

\begin{document}


\title{Collapse of rotating very massive stellar cores leading to a black hole and a massive disk as a source of gravitational waves}

\author[0000-0002-4979-5671]{Masaru Shibata}
\affiliation{Max-Planck-Institut f\"ur Gravitationsphysik (Albert-Einstein-Institut), Am M\"uhlenberg 1, D-14476 Potsdam-Golm, Germany}
\affiliation{Center for Gravitational Physics and Quantum Information, Yukawa Institute for Theoretical Physics, Kyoto University, Kyoto, 606-8502, Japan}

\author[0000-0001-6467-4969]{Sho Fujibayashi}
\affiliation{Frontier Research Institute for Interdisciplinary Sciences, Tohoku University, Aramaki aza Aoba 6-3, Aoba-ku, Sendai 980-8578, Japan}
\affiliation{Astronomical Institute, Graduate School of Science, Tohoku University, Sendai 980-8578, Japan}
\affiliation{Max-Planck-Institut f\"ur Gravitationsphysik (Albert-Einstein-Institut), Am M\"uhlenberg 1, D-14476 Potsdam-Golm, Germany}




\date{\today}


\begin{abstract}
We derive models of rotating very massive stellar cores with mass $\approx 10^2$--$10^4M_\odot$ which are marginally stable to the pair-unstable collapse, assuming that the core is isentropic and composed primarily of oxygen. It is shown that the cores with mass $\alt 10^3M_\odot$ can form a massive disk with the mass more than $10\%$ of the core mass around the formed black hole if the core is rotating with more than 30\% of the Keplerian limit. We also indicate that the formation of rapidly spinning massive black holes such as the black holes of GW231123 naturally accompanies the massive disk formation. By using the result of our previous study which showed that the massive disk is unstable to the non-axisymmetric deformation, we predict the amplitude and frequency of gravitational waves and show that the collapse of rotating very massive stellar cores can be a promising source of gravitational waves for Einstein Telescope. The detection of such gravitational waves will provide us with important information about a formation process of intermediate mass black holes.
\end{abstract} 


\section{Introduction}

Metal-poor very massive stars with initial mass larger than $\sim 260M_\odot$ are believed to collapse into a black hole after the onset of the electron-positron pair creation instability (referred to as pair instability in the following:~\citealt{2001ApJ...550..372F, Heger:2001cd}). Such stars evolve through hydrogen and helium burning, forming a core composed primarily of oxygen \citep{Bond:1984sn, 1996snih.book.....A}, and eventually become unstable to the pair instability, collapsing into a black hole. Such very massive stars are a plausible origin for intermediate-mass black holes. 

In the presence of a certain amount of angular momentum, the fate of the collapse could be rich because a disk should be formed around the formed black hole~\citep{Uchida:2018ago}. In the presence of a massive disk, the remnant can be a source of energetic phenomena such as a gamma-ray burst and an energetic supernova~(e.g., \citealt{2001ApJ...550..372F, Heger:2001cd, Uchida:2018ago, Siegel:2021ptt, 2025arXiv250815887G}). Such a system can be also a strong gravitational-wave emitter if the remnant disk is massive enough, i.e., the disk mass is larger than $\sim 10\%$ of the black hole mass~\citep{2011PhRvD..83d3007K, Kiuchi:2011re, Shibata:2021sau}. In this paper we will point out that collapse of very massive stellar cores with mass $\alt 10^3M_\odot$ can form a massive disk around the formed black hole if the surface of the core rotates with the angular velocity which is $\agt 30\%$ of the Keplerian one, and as a result, the remnant can be a strong emitter of gravitational waves in particular for Einstein Telescope~\citep{Hild:2010id} if the mass of the formed black holes is $\alt 200M_\odot$

In a recent paper~\citep{LIGOScientific:2025rsn}, a discovery of a high-mass binary black hole (GW231123) was reported. In this event, masses of two black holes are estimated to be $137^{+22}_{-17}M_\odot$ and $103^{+20}_{-52}M_\odot$ with 90\% credible intervals, and dimensionless spins for each black hole are quite high,  $0.90^{+0.10}_{-0.19}$ and $0.80^{+0.20}_{-0.51}$, respectively. If the black holes of GW231123 were formed from a collapse of a massive, rapidly rotating stellar core with mass $>140M_\odot$, it was likely to be evolved from a rotating very massive star with initial mass larger than $260M_\odot$ (see, e.g.,~\citealt{Takahashi2018, Tanikawa:2025fxw, Croon:2025gol, Stegmann:2025cja, Popa:2025dpz, Kiroglu:2025vqy} for formation scenarios). \citet{Uchida:2018ago} already provided a numerical model for the formation of such a heavy black hole using a progenitor model based on a stellar evolution calculation for a rapidly rotating metal poor star by~\citet{Takahashi2018} and a numerical-relativity simulation with relevant physical input (see also \citealt{2025arXiv250815887G} for a more simplified model). However, in the previous work, we studied only one particular model for the heavy black hole formation. One motivation of our series of work will be to systematically clarify the formation process of heavy black holes for a wide range of core mass $M\approx 10^2$--$10^4M_\odot$, for which the collapse is triggered by the pair instability~\citep{1971reas.book.....Z, Shibata:2024xsl}.
At the same time, the formation of black hole-massive disk systems can be a source of burst-type gravitational waves because the massive disk can be unstable against non-axisymmetric deformation. The primary purpose of this paper is to point out this possibility based on numerical results of~\citet{Shibata:2021sau}; if the black holes of GW231123 were formed from the collapse of a rotating very massive stellar core, the collapse itself emitted gravitational waves of a high amplitude. This implies that the formation process of such heavy black holes can be a target of future ground-based gravitational-wave detectors such as Einstein Telescope~\citep{Hild:2010id}. 

This paper is organized as follows: In Sec.~\ref{sec2} we briefly describe our setup for computing rotating very massive stellar cores in equilibrium. In Sec.~\ref{sec3} the properties of such stellar cores are summarized, paying particular attention to the rotational effect.  In Sec.~\ref{sec4} we predict the remnant of the collapse of rotating very massive stellar cores and show that the remnant can be a rapidly spinning black hole surrounded by a massive disk if the core is moderately rapidly rotating. In Sec.~\ref{sec5}, the predicted properties of gravitational waves emitted from unstable massive disks orbiting black holes with mass in the range of 50--$200M_\odot$ are summarized, indicating that the gravitational waves can be a source for Einstein Telescope~\citep{Hild:2010id}. Section~\ref{sec6} is devoted to a summary. 
Throughout this paper, $c$, $G$, and $k$ denote the speed of light, gravitational constant, and Boltzmann's constant, respectively. 

\section{Setup}\label{sec2}

\begin{figure*}[t]
\includegraphics[width=0.49\textwidth]{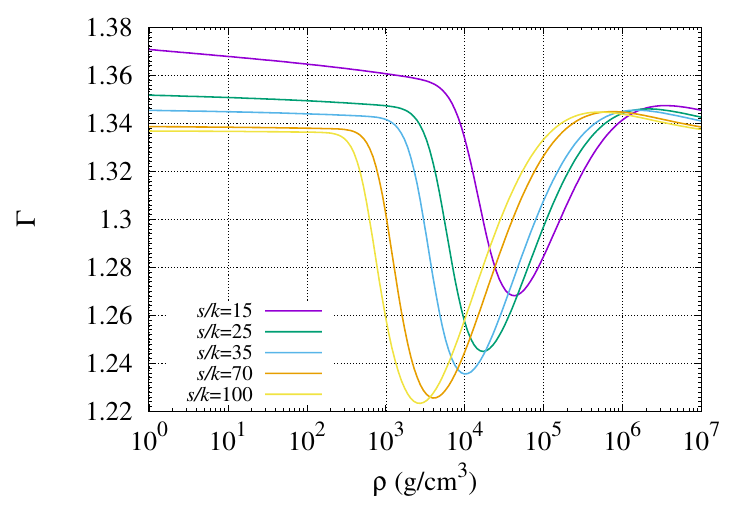}~~
\includegraphics[width=0.49\textwidth]{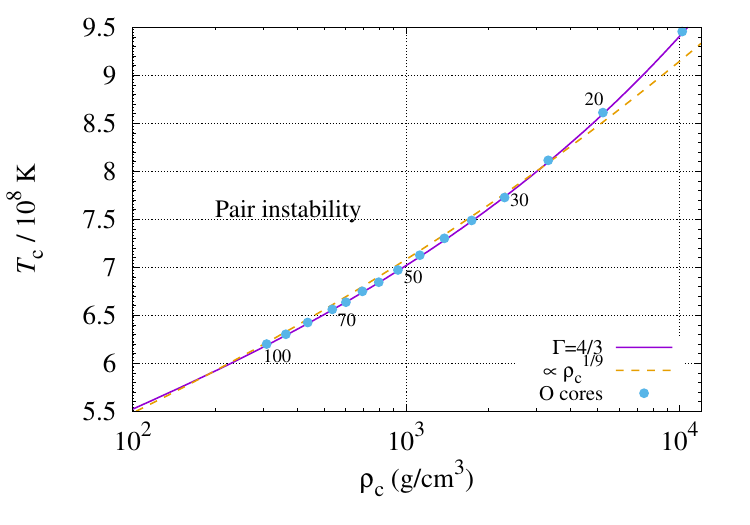}
\caption{Left: The adiabatic index $\Gamma$ as a function of the density $\rho$ for $s/k=15$, 25, 35, 70, and 100. Right: The solid curve denotes the curve of $\Gamma=4/3$ in the density-temperature plane. The dashed curve denotes a relation of $T\propto\rho^{1/9}$ as an approximation of the solid curve. The filled circles represent the physical information at the center of marginally-stable very massive stellar cores with $s/k=15$, 20, 25, 30, 35, 40, 45, 50, 55, 60, 65, 70, 80, 90, and 100.}
\label{fig0}
\end{figure*}

\begin{figure}[th]
\includegraphics[width=0.49\textwidth]{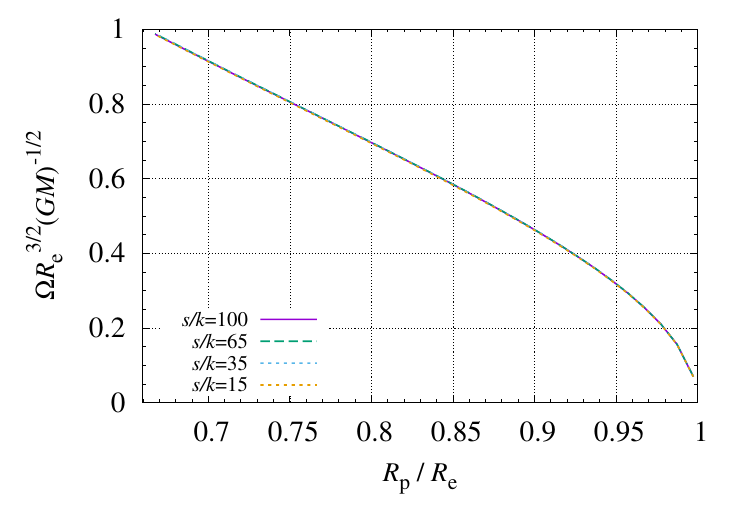}
\caption{The angular velocity $\Omega$ in units of $(GM/R_\mathrm{e}^3)^{1/2}$ as a function of $R_\mathrm{p}/R_\mathrm{e}$ for $s/k=15$, 35, 65, and 100. The 4 curves approximately overlap each other.}
\label{fig00}
\end{figure}

We numerically compute equilibrium states of rotating very massive stellar cores in general relativity which are marginally stable to the pair instability~\citep{1971reas.book.....Z} as the plausible initial condition for the pair-unstable collapse. According to stellar evolution calculations (e.g.,~\citealt{2016MNRAS.456.1320T, Takahashi2018}), the very massive stars become pair-unstable at a late phase of the evolution at which oxygen-carbon cores become sufficiently massive. In such a stage, the oxygen-carbon core is nearly isentropic and the mass fraction of oxygen is much larger than that of carbon. 
Thus, we employ a Timmes-Swesty equation of state~\citep{2000ApJS..126..501T} in the assumption that the stellar core is composed of fully ionized oxygens, electrons, positrons, and photons assuming uniform entropy per baryon, $s=$\,const. Specifically, $s/k$ is chosen to be larger than 13 
because for smaller values of $s/k$ the very massive star cores are likely to explode after the onset of the pair-instability (e.g.,~\citealt{1996snih.book.....A, 2001ApJ...550..372F, 2016MNRAS.456.1320T}). Note that \citet{2016MNRAS.456.1320T} reported the black hole formation from oxygen-carbon cores only with mass larger than $\approx 120M_\odot$. Thus, for $s/k \alt 14$ for which the oxygen core mass is $\alt 128M_\odot$ (see Sec.~\ref{sec3}), the oxygen core may explode after the onset of the collapse due to the pair instability. 

Rigid rotation is assumed following the stellar evolution results for rapidly rotating very massive stars by \citet{2016MNRAS.456.1320T, Takahashi2018}, which showed that the oxygen-carbon cores are broadly rigidly rotating with a weak degree of differential rotation. For example,~\citet{Takahashi2018} presented the end result for the evolution of a rapidly rotating very massive star with the initial mass $320M_\odot$. They found that at the critical density and temperature of the pair instability, the core is approximately composed of oxygen and carbon with the mass ratio 9:1 and its mass is $\approx 150M_\odot$ and is surrounded by an extended envelope composed of helium and hydrogen with its mass $\sim 140M_\odot$. The value of the entropy per baryon for the core is $\approx 15k$. \citet{Uchida:2018ago} performed an axisymmetric simulation in general relativity for the collapse of this stellar core and indeed showed that the remnant is a rapidly spinning black hole with the mass $\sim 130 M_\odot$ and the dimensionless spin of $\sim 0.8$, which is surrounded by a massive compact disk of mass $\sim 20M_\odot$.

Our aim is to numerically derive equilibrium stellar cores marginally stable against the pair instability. Thus, central values of the density and temperature are determined from the condition of $\Gamma=4/3$ for a given equation of state with $s=$const. For the chosen equations of state, the $\Gamma$ value steeply decreases far below $4/3$ with the increase of the density at such ranges of the density and temperature (see the left panel of Fig.~\ref{fig0}). On the other hand, for another range of the equation of state with lower density and temperature, the condition of $\Gamma > 4/3$ is always satisfied (see Fig.~\ref{fig0}). The criterion for the stability depends on compactness and rotational effect of the system~\citep{1978trs..book.....T, 1983bhwd.book.....S, Shibata:2016vxy}. However, these effects change the stability criterion of $\Gamma$ only by $O(10^{-3})$ in the current problem, and hence, we simply use $\Gamma=4/3$ as the critical curve for the stability. Note however that for rapidly rotating stellar cores near the mass shedding limit, the stabilization effect by the rotation can change the critical value of $\Gamma$ by $5 \times 10^{-3}$, and hence, the instability could happen for a more compact state than that for $\Gamma=4/3$.

The left panel of Fig.~\ref{fig0} shows $\Gamma$ as a function of the density for several values of $s/k$ in the range between 15 and 100. Irrespective of the $s/k$ values, $\Gamma$ is slightly larger than $4/3$ for low-density regions of $\rho \alt 10^{2.5}\,\mathrm{g/cm^3}$ for which the radiation pressure is the major pressure source. For the region of $\rho\approx 10^{2.5}$--$10^{5.5}\,\mathrm{g/cm^3}$, the value of $\Gamma$ becomes below $4/3$ because of the high temperature sufficient to create electron-positron pairs. For computing equilibrium states of very massive stellar cores, we choose the central density of $\rho_\mathrm{c}\approx 10^{2.5}$--$10^4\,\mathrm{g/cm^3}$ for which the $\Gamma$ value becomes $\approx 4/3$. It is worthy to note that the temperature of $\Gamma=4/3$ is in a narrow range as $6.2$--$9.5\times 10^8$\,K for $\rho \approx 10^{2.5}$--$10^4\,\mathrm{g/cm^3}$ (see the right panel of Fig.~\ref{fig0}). 

The solid curve in the right panel of Fig.~\ref{fig0} shows the relation between the density and temperature for $\Gamma=4/3$. At each point along the curve, the values of $s/k$ are different. For the larger density the $s/k$
value is smaller: The filled circles show the points of the representative values of $s/k=15$, 20, 25, 30, 35, 40, 45, 50, 55, 60, 65, 70, 80, 90, and 100. The dashed curve denotes a power-law fitting formula of the solid curve. We find that it is well represented by $T \propto \rho^{1/9}$ or $\rho \propto T^9$ for the range of $s/k$ in which we are interested.

\begin{figure*}[t]
\includegraphics[width=0.49\textwidth]{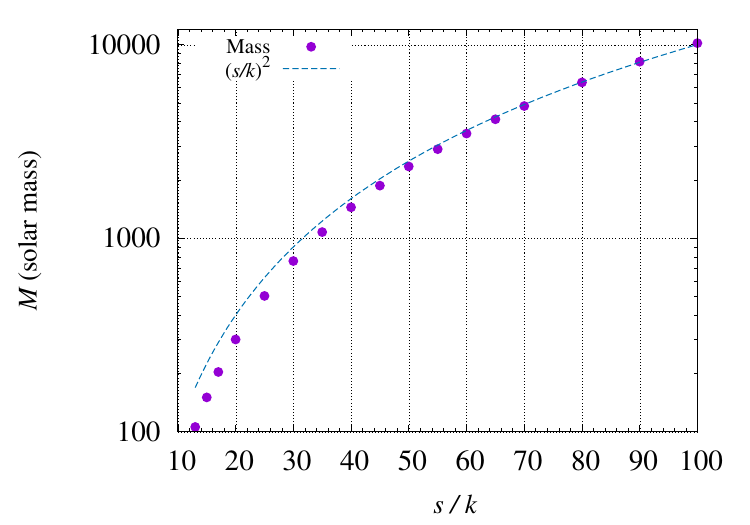}~~
\includegraphics[width=0.49\textwidth]{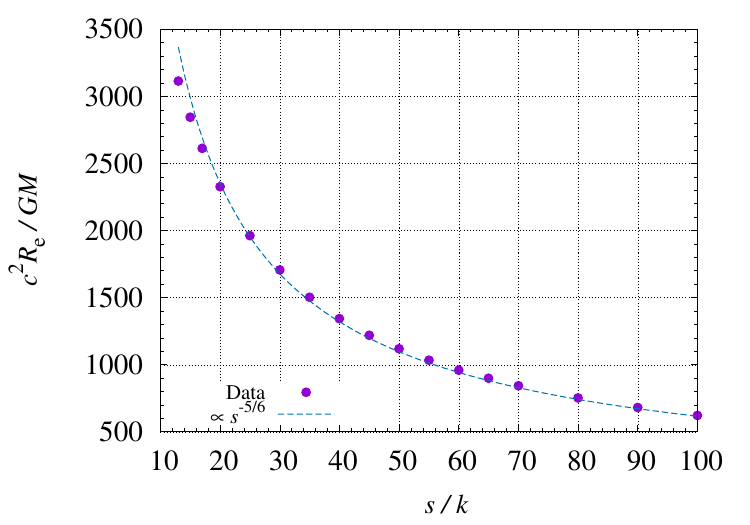}
\caption{$M$ (left) and $c^2R_\mathrm{e}/(GM)$ (right) of very massive stellar cores as functions of $s/k$ with axial ratios, $R_\mathrm{p}/R_\mathrm{e}=0.9975$. The dashed curves are $(s/k)^2$ (left) and $\propto s^{-5/6}$ (right).}
\label{fig1}
\end{figure*}

\begin{figure*}[t]
\includegraphics[width=0.49\textwidth]{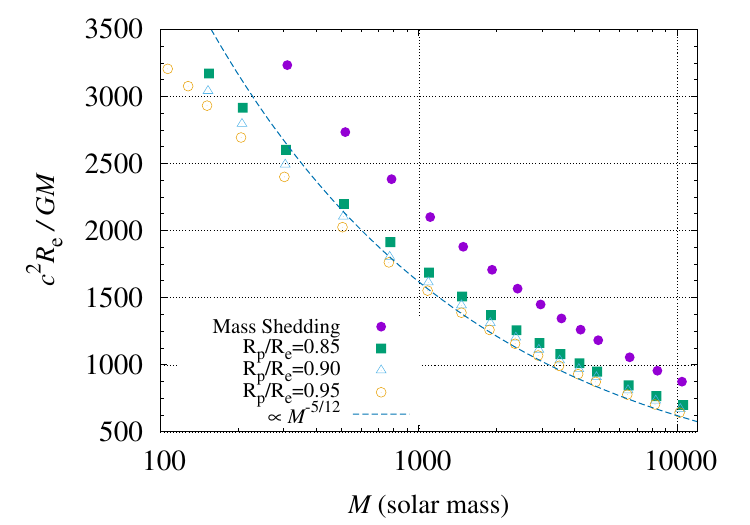}~~
\includegraphics[width=0.49\textwidth]{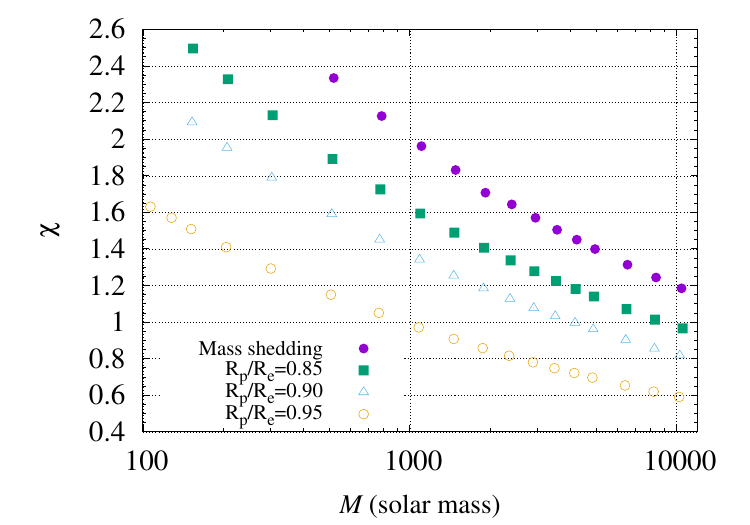}
\caption{The equatorial radius in units of $GM/c^2$ (left) and the dimensionless angular momentum $\chi=cJ/(GM^2)$ (right) with constant values of $R_\mathrm{p}/R_\mathrm{e}=2/3$, 0.85, 0.90, and $0.95$ as functions of the mass. The dashed curve in the left panel denotes $\propto M^{-5/12}$. For $R_\mathrm{p}R_\mathrm{e}=2/3$ the stellar cores are close to the mass shedding limit for which the angular velocity at the equatorial surface is $\approx 0.99 \sqrt{GM/R_\mathrm{e}^3}$.} 
\label{fig2}
\end{figure*}

Each model of the rotating stellar cores is determined by giving the axial ratio $R_\mathrm{p}/R_\mathrm{e}$ where $R_\mathrm{p}$ and $R_\mathrm{e}$ denote the polar and equatorial axial lengths, respectively. The numerical method is essentially the same as in our previous papers (e.g., \citealt{Shibata:2024xsl}). Because the stellar cores considered in this paper are not very compact (compactness defined by $GM/(c^2R_\mathrm{e})$ is smaller than $\sim 10^{-3}$), the coordinate axial lengths are in agreement with the proper length within $\sim 0.1\%$ error. Figure~\ref{fig00} shows the relation between the angular velocity $\Omega$ in units of the Keplerian one $\Omega_\mathrm{Kep}=(GM/R_\mathrm{e}^3)^{1/2}$ as a function of the axial ratio $R_\mathrm{p}/R_\mathrm{e}$ for $s/k=15$, 35, 65, and 100. We find that the relation depends only weakly on $s/k$ (i.e., mass $M$). In the following we pay attention to the cases with $R_\mathrm{p}/R_\mathrm{e}=2/3$, 0.85, 0.90, and 0.95, for which $\Omega/\Omega_\mathrm{Kep}$ is $\approx 0.99$, 0.59, 0.46, and 0.32, respectively.

\section{Properties of rotating very massive stellar cores}\label{sec3}

Before going ahead, we briefly discuss general properties for the cores of very massive stars, which are marginally stable against pair instability. As summarized, e.g., in~\citet{Bond:1984sn, Shibata:2024xsl}, the mass of very massive stellar cores $M$ is approximately proportional to $(s/k)^2$ for a given chemical composition because their adiabatic indices are close to 4/3. As shown in the left panel of Fig.~\ref{fig1}, this is indeed approximately satisfied. Also, the photon radiation always governs the equation of state, and hence, the entropy per baryon is approximately proportional to $T_\mathrm{c}^3/\rho_\mathrm{c}$ where $\rho_\mathrm{c}$ and $T_\mathrm{c}$ denote the central density and central temperature, respectively. Because $\rho_\mathrm{c} \propto T_\mathrm{c}^9$ is approximately satisfied at the onset of the pair instability, we approximately get $\rho_\mathrm{c} \propto s^{-3/2}$ and $T_\mathrm{c}\propto s^{-1/6}$.

For such stellar cores, the stellar radius $R$ in units of the gravitational radius $GM/c^2$ is written approximately as
\begin{eqnarray}
{c^2 R \over GM} &\propto& \rho_\mathrm{c}^{-1/3} M^{-2/3}
\nonumber \\
&\propto& s^{1/2}M^{-2/3} \propto s^{-5/6}\propto M^{-5/12},
\end{eqnarray}
where we used $\rho_c R^3 \propto M$ and considered spherical cores for simplicity. Therefore, the cores of very massive stars marginally stable against pair instability are less compact for the smaller core mass (i.e., smaller values of $s/k$). As shown in the right panel of Fig.~\ref{fig1} and the left panel of Fig.~\ref{fig2}, these proportionalities are indeed approximately satisfied.  

Assuming the rigid rotation of the core, the maximum angular velocity $\Omega$ is determined by the mass-shedding limit, i.e., 
\begin{eqnarray}
\Omega \leq \Omega_\mathrm{Kep}=\sqrt{{GM \over R_\mathrm{e}^3}}.
\end{eqnarray}
Since the angular momentum is approximately proportional to $MR_\mathrm{e}^2\Omega$, the maximum value of the dimensionless angular momentum defined by $\chi=cJ/(GM^2)$ is approximately proportional to $(R_\mathrm{e}/M)^{1/2}$. This implies that the value of $\chi$ can be larger for smaller masses, as shown in the right panel of Fig.~\ref{fig2}. This suggests that for smaller-mass cores, more rapidly spinning black holes can be formed with heavier disks surrounding the black hole. In the next section, we will show that this is indeed the case. 

\section{Predicted outcome of the collapse}\label{sec4}

\begin{figure*}[t]
\includegraphics[width=0.49\textwidth]{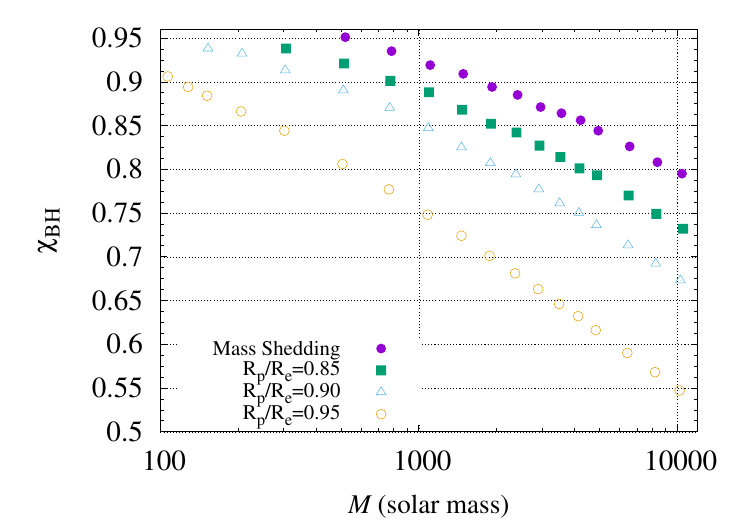}~~
\includegraphics[width=0.49\textwidth]{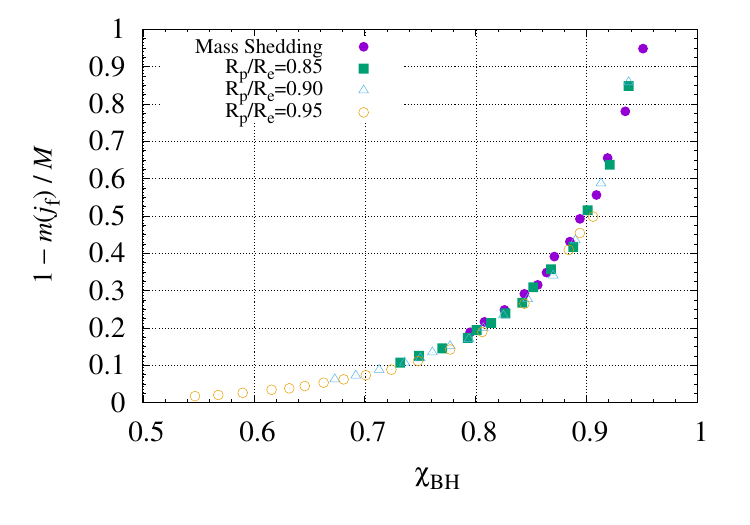}
\caption{Left: Predicted dimensionless spin of the hypothetically formed black hole, $\chi_\mathrm{BH}$, as a function of the core mass with fixed values of $R_\mathrm{p}/R_\mathrm{e}=2/3$, 0.85, 0.90, and 0.95. Right: The same as the left panel, but for the predicted mass fraction of the matter located outside the black hole as a function of $\chi_\mathrm{BH}$.
} 
\label{fig3}
\end{figure*}

\begin{figure*}[t]
\includegraphics[width=0.49\textwidth]{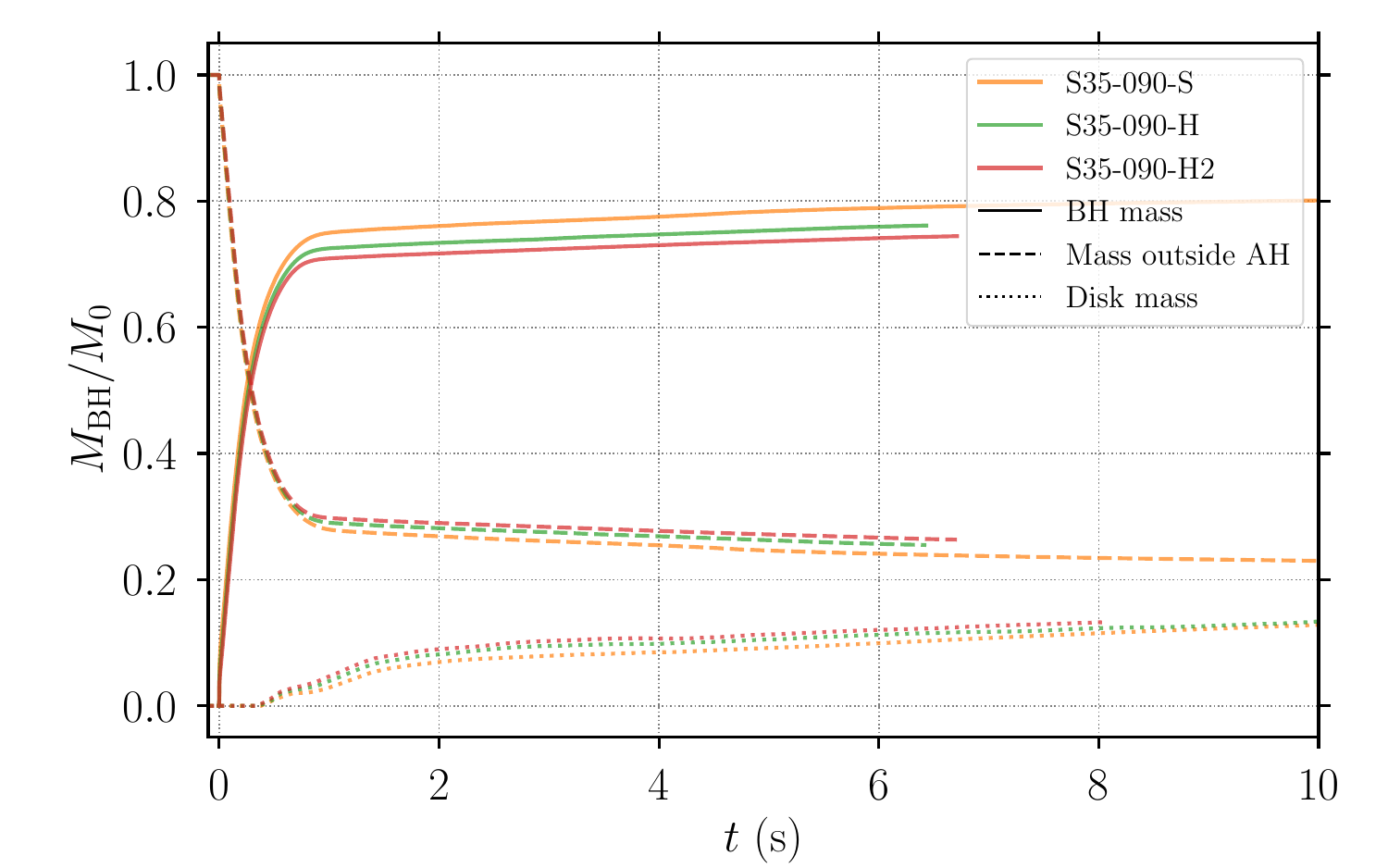}~~
\includegraphics[width=0.49\textwidth]{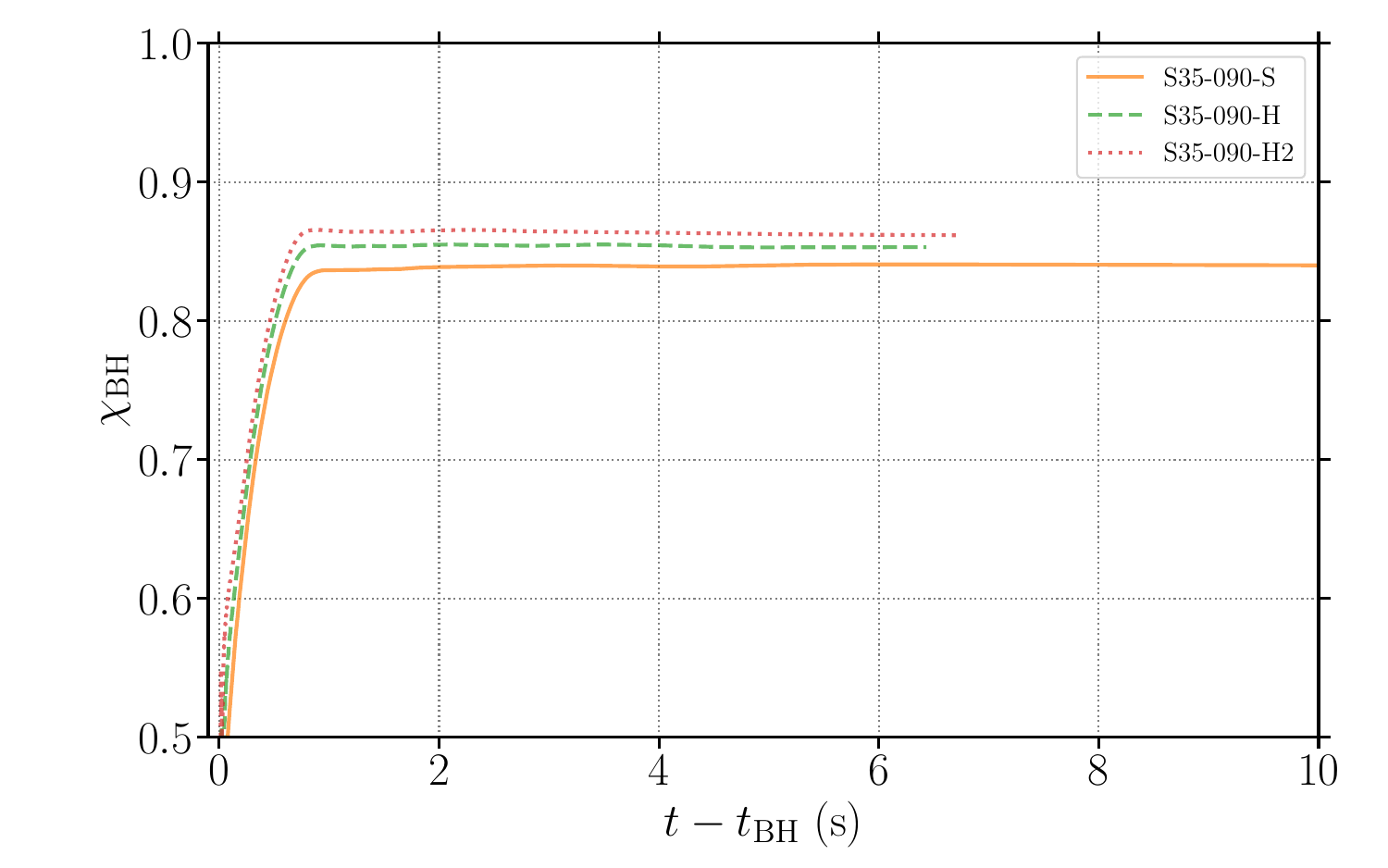}
\caption{Left: Evolution of masses of the black hole, of the matter located outside the black hole, and of the disk approximately defined as functions of time measured after the formation of the black hole, in units of the initial mass. Right: Evolution of the dimensionless spin of the black hole. For both panels, the results for $R_\mathrm{p}/R_\mathrm{e}=0.90$ are plotted. $t_\mathrm{BH}$ denotes the time at the formation of the black hole. 
} 
\label{fig4}
\end{figure*}

\begin{figure*}[t]
\hspace{0.0cm}\includegraphics[width=0.5\textwidth]{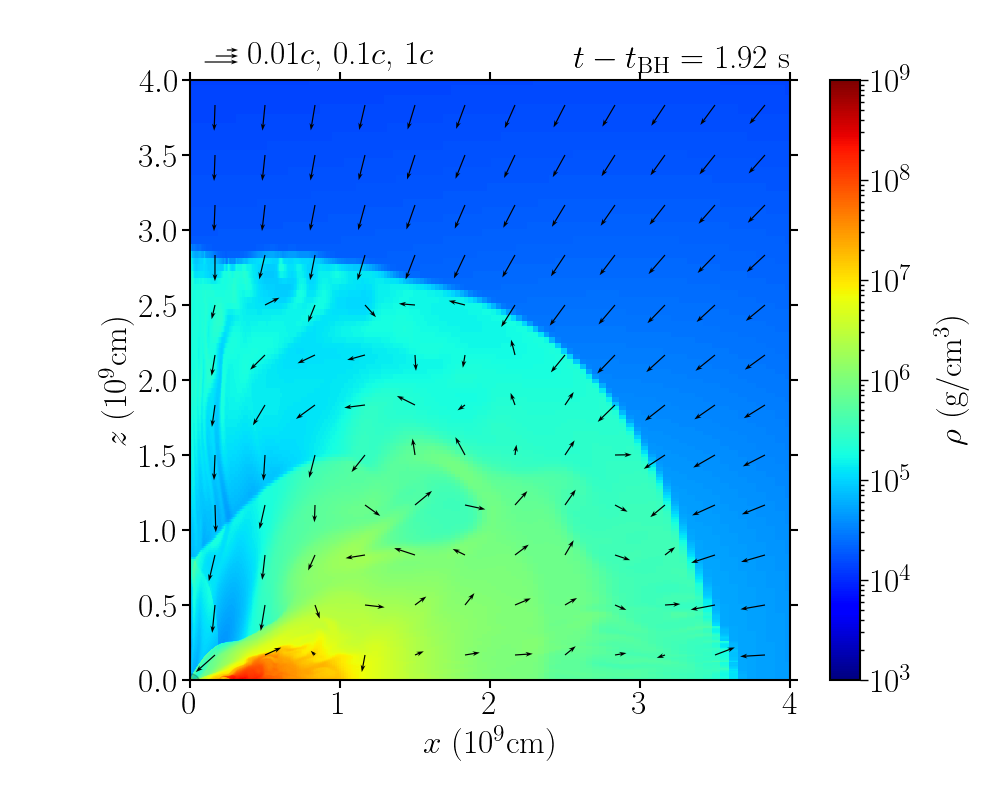}
\hspace{-0.0cm}\includegraphics[width=0.5\textwidth]{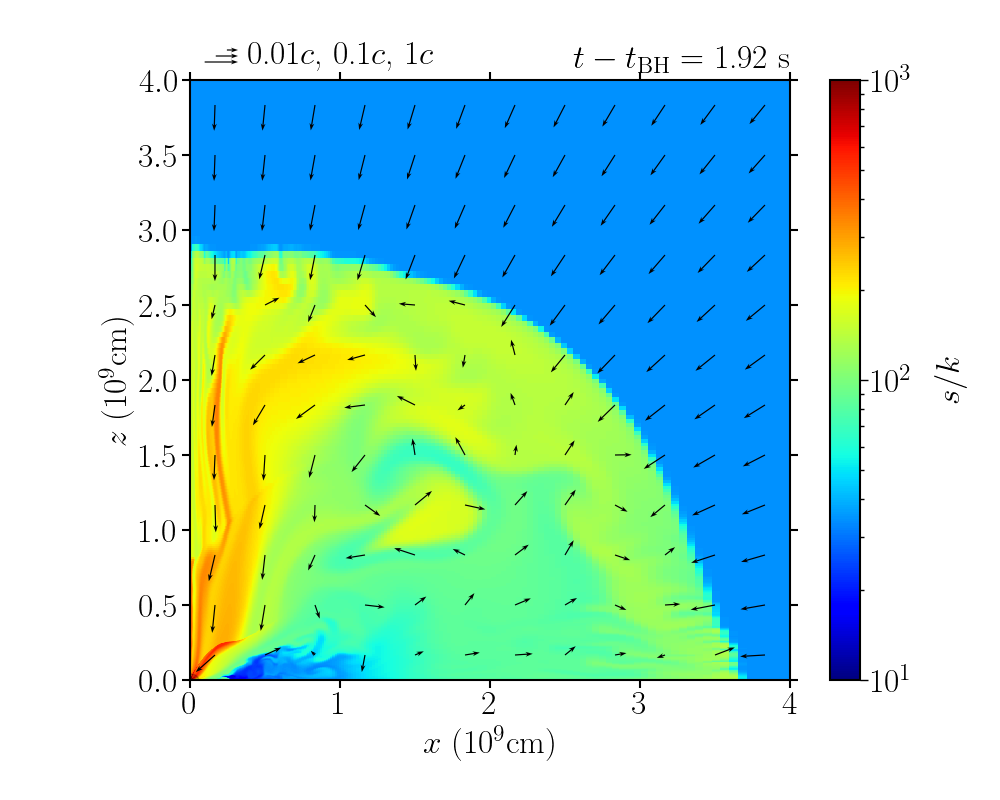}
\\
\hspace{0.0cm}\includegraphics[width=0.5\textwidth]{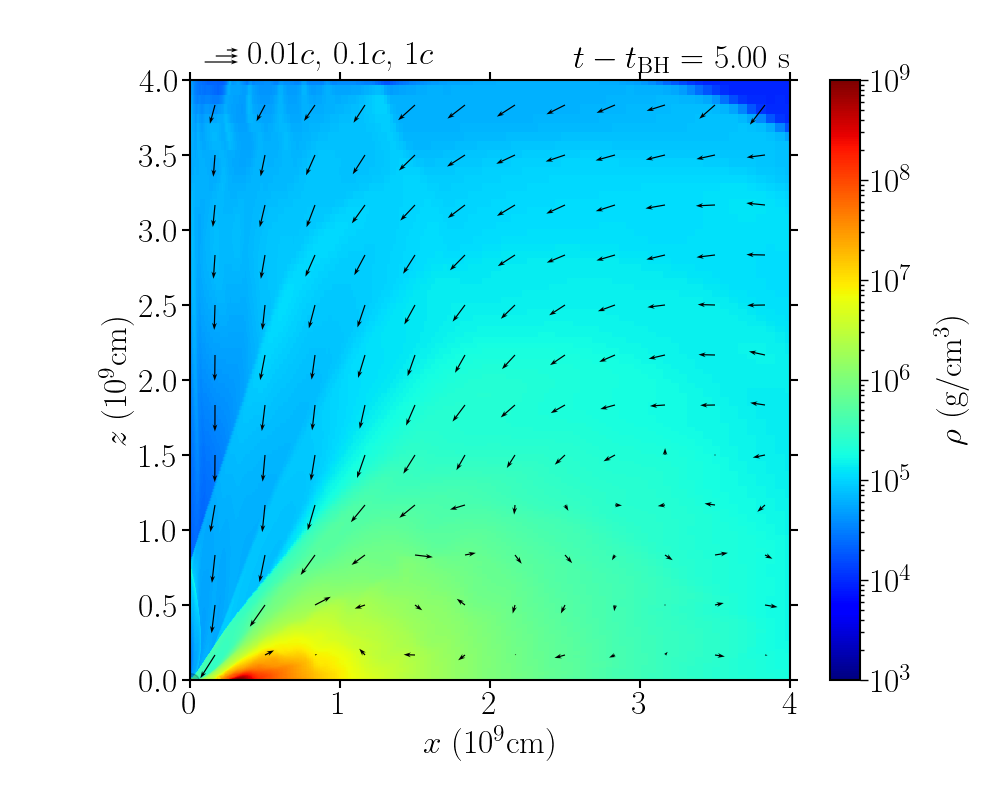}
\hspace{-0.0cm}\includegraphics[width=0.5\textwidth]{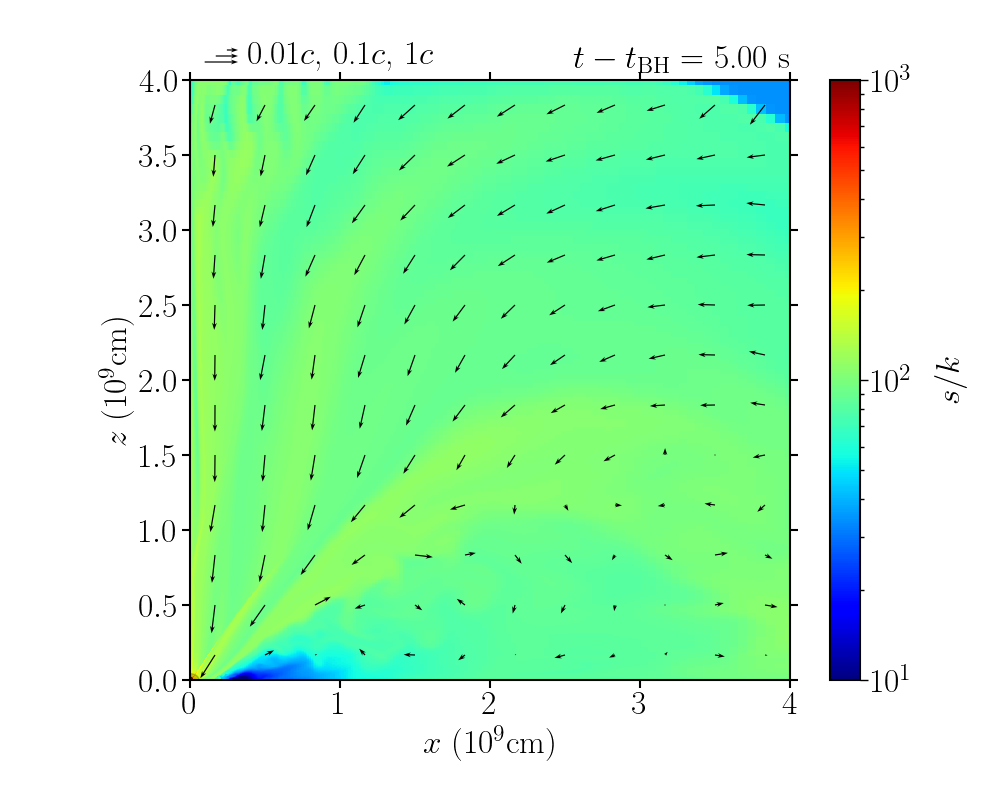}
\caption{Configuration of the rest-mass density (left) and entropy per baryon (right) of the disk at $t-t_\mathrm{BH} \approx 2$\,s (upper panel) and 5\,s (lower panel) for the simulation of $s/k=35$, $R_\mathrm{p}/R_\mathrm{e}=0.90$, and $\Delta x=0.007GM/c^2$. 
} 
\label{fig5}
\end{figure*}

Using the same procedure as in~\citet{Shibata:2002br, Shibata:2004vs, Shibata:2016vxy}, we predict the outcome of the collapse of rotating very massive stellar cores assuming that during the collapse leading to the formation of a black hole and a disk, the angular momentum transport plays a negligible role. As we already showed in the previous work~\citep{Uchida:2018ago}, the formation of the black hole and disk is achieved in the dynamical timescale of the system; for the core collapse of very massive stars, the core does not experience an appreciable bounce until the formation of the black hole (no formation of a proto-neutron star). Therefore, it is reasonable to assume the negligible angular momentum transport until the formation of the black hole-disk system. We expect that the angular momentum transport resulting from the magnetohydrodynamics effect will play an important role after the formation of the disk, but the relevant timescale is likely to be longer than the dynamical timescale with which the black hole-disk systems are formed, unless the viscosity or magnetic-field strength is extremely high (see, e.g., \citealt{Fujibayashi:2023oyt, Shibata:2023tho}).

Under the assumption that the angular momentum transport is negligible, the specific angular momentum, $j=h u_\varphi$, for each fluid element is conserved during the evolution of the system. Here, $h$ and $u_\varphi$ denote the specific enthalpy and azimuthal component of the four velocity with the lower subscript. We then calculate the mass and angular momentum of fluid elements in the oxygen cores with the specific angular momentum lower than a value $j$, denoted by $m(j)$ and $J(j)$, which satisfy $dm(j)/dj >0$ and $dJ(j)/dj>0$ for rigidly rotating objects. Numerical integration for $m(j)$ and $J(j)$ is performed for discrete values of $j$, and hence, a small numerical error appears in the predicted black hole mass and spin, although this does not change our conclusion. Here, the specific angular momentum increases with the increase of the cylindrical radius for $\Omega=$const., and thus, we assume that the fluid elements with smaller values of $j$ collapse into inner region earlier; $m(j)$ and $J(j)$ increase reflecting the distribution of $j$. Then it is natural to assume that the mass and angular momentum of a formed black hole increase with time, reflecting the distribution of $j$. 

Assuming that $m(j)$ and $J(j)$ are instantaneous values of the mass and angular momentum of a growing black hole, we can then derive the specific angular momentum of innermost stable circular orbits of the given black hole~\citep{Bardeen:1972fi}, $j_\mathrm{ISCO}(j)$, which is also a function of $j$. Note that the matter outside the black hole is assumed to have the specific angular momentum larger than $j$. Then, if $j_\mathrm{ISCO}$ is larger than $j$, we may consider that further accretion onto the black hole continues. On the other hand, if the condition of $j_\mathrm{ISCO} \leq j$ is satisfied, further accretion is prohibited and the matter should remain outside the black hole with the mass and angular momentum $m(j_\mathrm{f})$ and $J(j_\mathrm{f})$ where $j_\mathrm{f}$ denotes the value of $j$ which satisfies $j_\mathrm{ISCO}(j)=j$. We regard them as the predicted values of the black hole mass and angular momentum. The difference of the total mass $M$ and $m_\mathrm{f}$ can be regarded as the possibly maximum mass of a disk surrounding the formed black hole. We note that infalling matter of a non-circular orbit with $j>j_\mathrm{ISCO}$ may fall into the black hole in reality, and thus, the black hole mass can be slightly larger than the predicted value (see below).

Figure~\ref{fig3} shows the predicted black-hole dimensionless spin defined by $\chi_\mathrm{BH}=cJ(j_\mathrm{f})/(Gm(j_\mathrm{f})^2)$ as a function of the initial mass $M$ (left panel) and the maximum mass fraction that could be a disk defined by $1-m(j_\mathrm{f})/M$ (right panel) as a function of $\chi_\mathrm{BH}$. We plot the results for a variety of $s/k$ ($M\approx 10^2$--$10^4M_\odot$) with $R_\mathrm{p}/R_\mathrm{e}=2/3$, 0.85, $0.90$, and $0.95$. For $R_\mathrm{p}/R_\mathrm{e}=2/3$, the stellar core is approximately at the mass shedding limit. We note that $M-m(j_\mathrm{f})$ denotes the maximum mass that could form a disk. As shown below, the disk mass fraction is smaller than this during the evolution of the system, because the matter is still falling toward the center in the early stages.

The left panel of Fig.~\ref{fig3} shows that the black-hole dimensionless spin is likely to be high $\agt 0.8$ for $M\alt 10^3M_\odot$ even for moderately rapidly rotating cores with $R_\mathrm{p}/R_\mathrm{e}\alt 0.90$. In particular, for $M\alt 300M_\odot$, the dimensionless spin can be larger than 0.85 even for the stellar core with $R_\mathrm{p}/R_\mathrm{e}=0.95$ for which $\Omega/\Omega_\mathrm{Kep}$ is $\approx 32\%$. Such cores can be a good progenitor model of the black holes of GW231123. 

The right panel of Fig.~\ref{fig3} shows that for $\chi_\mathrm{BH}\agt 0.80$, 0.85, and 0.90, $\agt 20\%$, 30\%, and 50\% of the initial mass, respectively, could form a disk irrespective of the angular velocity of the stellar core. This indicates that the formation of a rapidly spinning black hole would accompany the formation of a massive disk, which will be subsequently unstable to non-axisymmetric deformation. It should be also noted that the black-hole mass can be much smaller than the core mass for the rapidly rotating core collapse. For example, a black hole with mass $\sim 50M_\odot$ may be temporarily formed from a core with mass $\agt 140M_\odot$, if the formed black hole is rapidly spinning (after the evolution of the formed disk by the non-axisymmetric instability and/or viscous processes, the black-hole mass would increase subsequently). 


To confirm that our analysis is approximately valid, we perform axisymmetric simulations in general relativity. For this purpose, we picked up a stellar core with $s/k=35$ and $R_\mathrm{p}/R_\mathrm{e}=0.85$, 0.90, and 0.95. For these models, the core mass is $\approx 1.10$, $1.09$, and $1.08 \times 10^3M_\odot$. For the simulation, we use a code recently developed for a study of the collapse of supermassive stars~\citep{Fujibayashi:2024vnb}, for which the Timmes-Swesty equation of state is employed. We include the neutrino emission cooling by pair processes adopting the fitting formula in \cite{1996ApJS..102..411I}. We further implement a nuclear reaction network as in the manner of \citet{Uchida:2018ago,1999ApJS..124..241T}, although for such massive stellar cores, the effect of the nuclear burning in the formation of a black hole and a disk is minor. Simulations are performed on a non-uniform grid structure of cylindrical coordinates of $\varpi$-$z$ with the three grid resolutions of $\Delta x/M=0.015$, $0.010$, and $0.007$, where $x$ denotes $\varpi$ or $z$. The grid spacing is uniform for $x \alt GM/c^2$ and increased as $dx_{i+1}=1.017 dx_i$ for $x \agt GM/c^2$ where $x_i$ denotes the $i$-th grid point.

Figure~\ref{fig4} shows the evolution of masses of the black hole, of the matter located outside the black hole, and of the disk approximately defined as a function of the time measured after the formation of the black hole in units of the initial mass (left) and the evolution of the dimensionless spin of the black hole for $R_\mathrm{p}/R_\mathrm{e}=0.90$. Here, the disk is defined as a region in which the radial velocity is smaller than the rotational velocity. It is found that the black hole mass and dimensionless spin increase with time toward approximate constants. The relaxed values of the black hole mass and dimensionless spin decrease and increase, respectively, with the improvement of the grid resolution and the convergence is not fully achieved in our choice of the grid resolution. However, it is reasonable to conclude that the masses of the black hole and the matter located outside the black hole at $t-t_\mathrm{BH}=2$\,s are $\approx 0.7M$ and $0.3M$ (i.e., $M_\mathrm{disk}/M_\mathrm{BH}$ can be $\sim 0.4$), respectively, and the dimensionless spin is $\approx 0.86$. These values agree approximately with those predicted from the initial configuration (compare with the right panel of Fig.~\ref{fig3}).

Figure~\ref{fig5} shows the configuration of the rest-mass density (left) and entropy per baryon (right) of the disk at $t-t_\mathrm{BH}\approx 2$\,s and 5\,s. It is found that the density maximum is located at $\approx 3$--$4GM_\mathrm{BH}/c^2$ and high-density (and low-entropy) region is confined in the region of $\varpi \alt 15GM_\mathrm{BH}/c^2$ (note that $M_\mathrm{BH}\approx 8\times 10^2M_\odot$ and thus $GM_\mathrm{BH}/c^2 \approx 1.2 \times 10^8$\,cm). This implies that a compact and massive disk is indeed formed immediately after the black hole formation, and in the subsequent time evolution, an extended region of the disk grows. In the high-density part of the disk, the values of $s/k$ are approximately the same as the initial value, $s/k=35$, besides a decrease due to neutrino cooling. This indicates that if the initial values of $s/k$ are approximately preserved, a compact disk would be formed. 

As Fig.~\ref{fig4} shows, the mass of the disk increases with time gradually due to the matter infall from the outer region. As previous studies showed~\citep{Kiuchi:2011re, Shibata:2021sau}, the more massive (and more compact) disk can be more unstable to non-axisymmetric deformation. Hence, it is not clear at what time the most intense instability sets in during the evolution of the disk. To fully understand the mechanism of the non-axisymmetric instability, we need a three-dimensional simulation, which we plan to perform in future work. However, the present result clearly shows that a massive disk is formed, and therefore, gravitational waves associated with the non-axisymmetric deformation of the massive disk should be emitted after the collapse of rotating very massive stellar cores to a black hole.

We performed similar simulations for $R_\mathrm{p}/R_\mathrm{e}=0.85$ and 0.95. For these models, the masses of the black hole and matter located outside the black hole and the dimensionless spin of the black hole approach $\approx 0.6M$, $0.4M$,  
and $0.91$ for $R_\mathrm{p}/R_\mathrm{e}=0.85$ and $\approx 0.9M$, $0.1M$, and $0.75$ for $R_\mathrm{p}/R_\mathrm{e}=0.95$, respectively. The dimensionless spins of the black holes are broadly consistent with those predicted from the initial data (compare with the right panel of Fig.~\ref{fig3}). The value of $M-m(j_\mathrm{f})$ for the rapidly spinning case with $R_\mathrm{p}/R_\mathrm{e}=0.85$ is slightly smaller than the prediction from the initial data, but it is clearly shown that the high initial rotation can enhance the fraction of the matter located outside the black hole. 

\section{Discussion}\label{sec5}

\citet{Shibata:2021sau} performed fully general relativistic simulations for massive disks with mass of $M_\mathrm{disk}=15$--$50M_\odot$ orbiting a spinning black hole with mass of $M_\mathrm{BH}\approx 50M_\odot$ and dimensionless spin of $\chi_\mathrm{BH}\approx 0.8$. They considered disks with the maximum density located at $R_\mathrm{peak}=7$--$9\,r_g$ where $r_g=GM_\mathrm{BH}/c^2$. They found that for all the cases, the disk is non-axisymmetrically unstable to the formation of a spiral arm and becomes a burst source of gravitational waves. They also showed that some of the waveforms are similar to the observed waveform of GW190521~\citep{LIGOScientific:2020iuh} (and also to that of GW231123;~\citealt{LIGOScientific:2025rsn}): the waveforms are composed primarily of a few burst waves with high amplitude and subsequent quasi-periodic waves with relatively low amplitude.

\begin{figure}[t]
\includegraphics[width=0.49\textwidth]{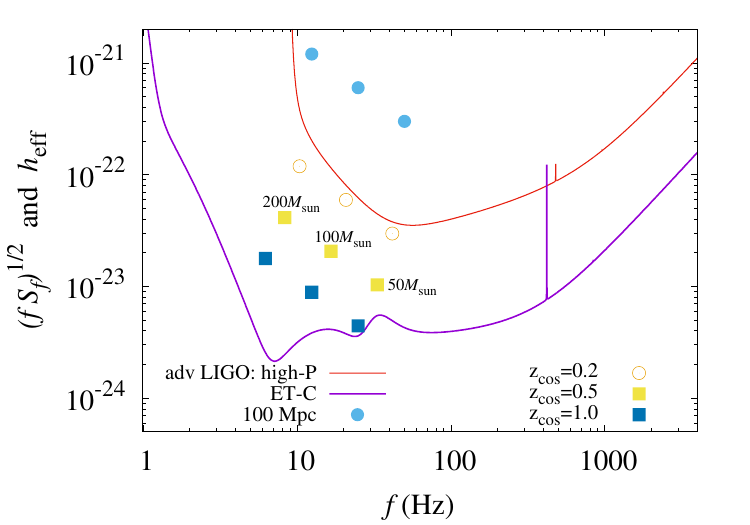}
\caption{Comparison of the predicted effective amplitude of gravitational waves for $M_\mathrm{BH}=200$, $100$, and $50M_\odot$ and designed sensitivity curves of the advanced LIGO and Einstein Telescope of the configuration-type C. We choose $\epsilon=0.2$, $M_\mathrm{disk}/M_\mathrm{BH}=0.4$, and $R_\mathrm{peak}/r_g=7$ for the plot of the effective amplitude. The luminosity distances are chosen to be $100$\,Mpc, and those of the cosmological redshift $z_\mathrm{cos}=0.2$, 0.5, and 1.0. The cosmological effects are taken into account to plot the effective amplitude and frequency.}
\label{fig6}
\end{figure}

The typical frequency of gravitational waves (in the source frame) we found in \cite{Shibata:2021sau} is written approximately as
\begin{eqnarray}
f_\mathrm{GW} \approx 0.8 {c \over \pi r_g}\left({R_\mathrm{peak} \over r_g}\right)^{-3/2},
\end{eqnarray}
and for $M_\mathrm{BH}=50M_\odot$ it is 40--60\,Hz. Here the approximate factor of 0.8 comes from the fact that the spiral arm is formed in an extended region of $r > R_\mathrm{peak}$. 
The maximum amplitude of gravitational waves emitted is written as
\begin{eqnarray}
h_\mathrm{eff}=\epsilon {GM_\mathrm{disk} r_g \over c^2D R_\mathrm{peak}}
&\approx& 3 \times 10^{-22} \left({\epsilon \over 0.2}\right)
\left({M_\mathrm{disk} \over 20M_\odot}\right)
\nonumber \\ &&\times 
\left({R_\mathrm{peak} \over 7r_g}\right)^{-1} 
\left({D \over 100\,\mathrm{Mpc}}\right)^{-1}, 
\end{eqnarray}
where $D$ denotes the luminosity distance to the source, $\epsilon$ is approximately in the range between 0.1 and 0.3, and the gravitational wave amplitude is consistent with the prediction by the quadrupole formula. We note that the value of $\epsilon$ depends on the compactness (or equivalently the entropy per baryon) of the disk. 

We here focus on the cases that the disk mass is 30--60\% of the black hole mass, and extrapolate our previous results for higher black hole masses with $M_\mathrm{BH}\agt 100M_\odot$. For this case, the gravitational-wave frequency becomes lower as
\begin{eqnarray}
f_\mathrm{GW} \approx 28 \mathrm{Hz}\left({M_\mathrm{BH} \over 100M_\odot}\right)^{-1}\left({R_\mathrm{peak} \over 7r_g}\right)^{-3/2}. 
\end{eqnarray}
On the other hand, the disk mass is larger for a fixed ratio of $M_\mathrm{disk}/M_\mathrm{BH}$. 

Figure~\ref{fig6} plots the predicted effective amplitudes and frequencies of gravitational waves for $\epsilon=0.2$, $M_\mathrm{disk}/M_\mathrm{BH}=0.4$, and $R_\mathrm{peak}/r_g=7$, and compares them with the designed sensitivity curves of advanced LIGO and Einstein Telescope of the configuration-type C~\citep{Hild:2010id}
(see also \url{https://www.et-gw.eu/index.php/etsensitivities}). This shows that with the current detectors, the events for $M_\mathrm{BH}=100$--$200M_\odot$ can be detected only when the distance to the source is less than $\sim 100$--200\,Mpc. On the other hand, by Einstein Telescope for which the sensitivity in the lower frequency band $\alt 30$\,Hz is much better than that of the advanced LIGO, the events with $M_\mathrm{BH}=100$--$200M_\odot$ may be detected up to $z_\mathrm{cos}\sim 1$. As the supernova rate increases with the increase of $z_\mathrm{cos}$ (see, e.g.,~\citealt{2025arXiv250810985P}) and the metallicity of massive stars should be lower for higher cosmological redshifts, it may be expected that the formation rate of the very massive stars leading to black hole formation can be higher than that in the current universe. This suggests that the collapse of rotating very massive stellar cores at $z_\mathrm{cos}\alt 1$ can be a source of Einstein Telescope at lower frequency band of $f_\mathrm{GW}\sim 10$--20\,Hz; if the formation of massive black holes similar to the black holes of GW231123 is frequently formed for $z_\mathrm{cos}\alt 1$, the formation process itself is the promising source of Einstein Telescope.

Our numerical simulations for $s/k=35$ show that $R_\mathrm{peak}/r_g$ is appreciably smaller than 7. If this is also the case for smaller values of $s/k$ by which black holes of $M_\mathrm{BH}=50$--$200M_\odot$ are formed, the frequency and amplitude of gravitational waves considered here are higher and larger, respectively. We plan to systematically investigate this point in our future numerical work. 

One concern associated with the detection of this type of burst gravitational waves is that the waveforms can be mistaken for that of a binary black hole merger because two gravitational waveforms are similar for the highest amplitude part. An advantage for the gravitational collapse case is that this type of event is likely to be accompanied by electromagnetic emissions, which can be driven by a disk wind~\citep{Uchida:2018ago, Siegel:2021ptt, 2025arXiv250315729A} and/or a jet~\citep{2025arXiv250815887G}. Thus, the detection of some electromagnetic counterparts will be the key to confirming that the gravitational-wave event is associated with the gravitational collapse of a very massive star.

\section{Summary}\label{sec6}

We derived and analyzed equilibrium states of rotating oxygen cores that are marginally stable against pair instability. We showed that for relatively low-mass cores with $M\alt 10^3M_\odot$, the radius of the cores in units of the gravitational radius can be quite large $\agt 1500$, and thus, even in the presence of a moderately rapid rotation, the rotational effect at the formation of a black hole can be significant. Specifically, if the core rotates with $\Omega/\Omega_\mathrm{Kep}\agt 0.3$, the mass of the disk can be $\agt 30\%$ of the initial core mass; a black hole can be surrounded by a massive disk which is subsequently unstable to non-axisymmetric instability and can be a strong emitter of gravitational waves. 

Using our previous study on the evolution of unstable massive disks orbiting a black hole~\citep{Shibata:2021sau}, we estimated the effective amplitude and frequency of gravitational waves emitted from the massive disk assuming that the disk mass is 30--60\% of the black-hole mass. We indicated that the black hole-disk systems with $M_\mathrm{BH}\approx 100$--$200M_\odot$ formed for $z_\mathrm{cos}\alt 1$ can be a source of Einstein Telescope for the frequency band of $\sim 10$--20\,Hz. 

In the present paper, we used our previous results of \cite{Shibata:2021sau} to infer the properties of gravitational waves assuming that the scaling relation for the amplitude and frequency with respect to the black-hole mass is satisfied. Although this assumption is likely to be valid, it is desirable to perform a numerical simulation for more strict setup for the masses of the black hole and disk. More desirable is to perform a simulation from the stellar core collapse throughout the formation of a black hole-disk system, which is subsequently unstable for the emission of gravitational waves. We also plan to perform such a self-consistent simulation in subsequent work. Exploring the long-term evolution of the disk is also an important subject to determine the final mass and spin of the black hole. 

In \citet{Shibata:2021sau}, we pointed out that the waveforms of another massive binary black hole merger, GW190521 \citep{LIGOScientific:2020iuh}, could be mimicked by gravitational waves from black hole-massive disk systems for which the disk is deformed non-axisymmetrically. This could also be the case for GW231123; similar waveforms may be derived by a model of black hole-disk systems. It is interesting to pursue this possibility as well. 

\acknowledgements
We thank Koh Takahashi for the useful discussion and for sharing the data of stellar evolution models. We also thank Kenta Hotokezaka for the helpful conversation. 
Numerical computation was performed on the cluster Sakura and Momiji at the Max Planck Computing and Data Facility. This work was in part supported by Grant-in-Aid for Scientific Research (grant No.~23H04900) of Japanese MEXT/JSPS.

\bibliography{reference}

\begin{thebibliography}{}
\expandafter\ifx\csname natexlab\endcsname\relax\def\natexlab#1{#1}\fi
\providecommand{\url}[1]{\href{#1}{#1}}
\providecommand{\dodoi}[1]{doi:~\href{http://doi.org/#1}{\nolinkurl{#1}}}
\providecommand{\doeprint}[1]{\href{http://ascl.net/#1}{\nolinkurl{http://ascl.net/#1}}}
\providecommand{\doarXiv}[1]{\href{https://arxiv.org/abs/#1}{\nolinkurl{https://arxiv.org/abs/#1}}}

\bibitem[{Abac {et~al.}(2025)}]{LIGOScientific:2025rsn}
Abac, A.~G., {et~al.} 2025.
\newblock \doarXiv{2507.08219}

\bibitem[{Abbott {et~al.}(2020)}]{LIGOScientific:2020iuh}
Abbott, R., {et~al.} 2020, Phys. Rev. Lett., 125, 101102,
  \dodoi{10.1103/PhysRevLett.125.101102}

\bibitem[{{Agarwal} {et~al.}(2025){Agarwal}, {Siegel}, {Metzger}, \&
  {Nagele}}]{2025arXiv250315729A}
{Agarwal}, A., {Siegel}, D.~M., {Metzger}, B.~D., \& {Nagele}, C. 2025, arXiv
  e-prints, arXiv:2503.15729, \dodoi{10.48550/arXiv.2503.15729}

\bibitem[{{Arnett}(1996)}]{1996snih.book.....A}
{Arnett}, D. 1996, {Supernovae and Nucleosynthesis: An Investigation of the
  History of Matter from the Big Bang to the Present}

\bibitem[{Bardeen {et~al.}(1972)Bardeen, Press, \& Teukolsky}]{Bardeen:1972fi}
Bardeen, J.~M., Press, W.~H., \& Teukolsky, S.~A. 1972, Astrophys. J., 178,
  347, \dodoi{10.1086/151796}

\bibitem[{Bond {et~al.}(1984)Bond, Arnett, \& Carr}]{Bond:1984sn}
Bond, J.~R., Arnett, W.~D., \& Carr, B.~J. 1984, Astrophys. J., 280, 825,
  \dodoi{10.1086/162057}

\bibitem[{Croon {et~al.}(2025)Croon, Sakstein, \& Gerosa}]{Croon:2025gol}
Croon, D., Sakstein, J., \& Gerosa, D. 2025.
\newblock \doarXiv{2508.10088}

\bibitem[{{Fryer} {et~al.}(2001){Fryer}, {Woosley}, \&
  {Heger}}]{2001ApJ...550..372F}
{Fryer}, C.~L., {Woosley}, S.~E., \& {Heger}, A. 2001, \apj, 550, 372,
  \dodoi{10.1086/319719}

\bibitem[{Fujibayashi {et~al.}(2025)Fujibayashi, Jockel, Kawaguchi, Sekiguchi,
  \& Shibata}]{Fujibayashi:2024vnb}
Fujibayashi, S., Jockel, C., Kawaguchi, K., Sekiguchi, Y., \& Shibata, M. 2025,
  Astrophys. J., 981, 119, \dodoi{10.3847/1538-4357/adb0b8}

\bibitem[{Fujibayashi {et~al.}(2024)Fujibayashi, Lam, Shibata, \&
  Sekiguchi}]{Fujibayashi:2023oyt}
Fujibayashi, S., Lam, A. T.-L., Shibata, M., \& Sekiguchi, Y. 2024, Phys. Rev.
  D, 109, 023031, \dodoi{10.1103/PhysRevD.109.023031}

\bibitem[{{Gottlieb} {et~al.}(2025){Gottlieb}, {Metzger}, {Issa}, {Li},
  {Renzo}, \& {Isi}}]{2025arXiv250815887G}
{Gottlieb}, O., {Metzger}, B.~D., {Issa}, D., {et~al.} 2025, arXiv e-prints,
  arXiv:2508.15887, \dodoi{10.48550/arXiv.2508.15887}

\bibitem[{Heger \& Woosley(2002)}]{Heger:2001cd}
Heger, A., \& Woosley, S.~E. 2002, Astrophys. J., 567, 532,
  \dodoi{10.1086/338487}

\bibitem[{Hild {et~al.}(2011)}]{Hild:2010id}
Hild, S., {et~al.} 2011, Class. Quant. Grav., 28, 094013,
  \dodoi{10.1088/0264-9381/28/9/094013}

\bibitem[{{Itoh} {et~al.}(1996){Itoh}, {Hayashi}, {Nishikawa}, \&
  {Kohyama}}]{1996ApJS..102..411I}
{Itoh}, N., {Hayashi}, H., {Nishikawa}, A., \& {Kohyama}, Y. 1996, \apjs, 102,
  411, \dodoi{10.1086/192264}

\bibitem[{K{\i}ro{\u{g}}lu {et~al.}(2025)K{\i}ro{\u{g}}lu, Kremer, \&
  Rasio}]{Kiroglu:2025vqy}
K{\i}ro{\u{g}}lu, F., Kremer, K., \& Rasio, F.~A. 2025.
\newblock \doarXiv{2509.05415}

\bibitem[{Kiuchi {et~al.}(2011)Kiuchi, Shibata, Montero, \&
  Font}]{Kiuchi:2011re}
Kiuchi, K., Shibata, M., Montero, P.~J., \& Font, J.~A. 2011, Phys. Rev. Lett.,
  106, 251102, \dodoi{10.1103/PhysRevLett.106.251102}

\bibitem[{{Korobkin} {et~al.}(2011){Korobkin}, {Abdikamalov}, {Schnetter},
  {Stergioulas}, \& {Zink}}]{2011PhRvD..83d3007K}
{Korobkin}, O., {Abdikamalov}, E.~B., {Schnetter}, E., {Stergioulas}, N., \&
  {Zink}, B. 2011, \prd, 83, 043007, \dodoi{10.1103/PhysRevD.83.043007}

\bibitem[{{Pessi} {et~al.}(2025){Pessi}, {Desai}, {Prieto}, {Kochanek},
  {Shappee}, {Anderson}, {Beacom}, {Dong}, {Stanek}, \&
  {Thompson}}]{2025arXiv250810985P}
{Pessi}, T., {Desai}, D.~D., {Prieto}, J.~L., {et~al.} 2025, arXiv e-prints,
  arXiv:2508.10985, \dodoi{10.48550/arXiv.2508.10985}

\bibitem[{Popa \& de~Mink(2025)}]{Popa:2025dpz}
Popa, S.~A., \& de~Mink, S.~E. 2025.
\newblock \doarXiv{2509.00154}

\bibitem[{{Shapiro} \& {Teukolsky}(1983)}]{1983bhwd.book.....S}
{Shapiro}, S.~L., \& {Teukolsky}, S.~A. 1983, {Black holes, white dwarfs and
  neutron stars. The physics of compact objects}, \dodoi{10.1002/9783527617661}

\bibitem[{Shibata(2004)}]{Shibata:2004vs}
Shibata, M. 2004, Astrophys. J., 605, 350, \dodoi{10.1086/382234}

\bibitem[{Shibata {et~al.}(2025)Shibata, Fujibayashi, Jockel, \&
  Kawaguchi}]{Shibata:2024xsl}
Shibata, M., Fujibayashi, S., Jockel, C., \& Kawaguchi, K. 2025, Astrophys. J.,
  978, 58, \dodoi{10.3847/1538-4357/ad93a4}

\bibitem[{Shibata {et~al.}(2024)Shibata, Fujibayashi, Lam, Ioka, \&
  Sekiguchi}]{Shibata:2023tho}
Shibata, M., Fujibayashi, S., Lam, A. T.-L., Ioka, K., \& Sekiguchi, Y. 2024,
  Phys. Rev. D, 109, 043051, \dodoi{10.1103/PhysRevD.109.043051}

\bibitem[{Shibata {et~al.}(2021)Shibata, Kiuchi, Fujibayashi, \&
  Sekiguchi}]{Shibata:2021sau}
Shibata, M., Kiuchi, K., Fujibayashi, S., \& Sekiguchi, Y. 2021, Phys. Rev. D,
  103, 063037, \dodoi{10.1103/PhysRevD.103.063037}

\bibitem[{Shibata \& Shapiro(2002)}]{Shibata:2002br}
Shibata, M., \& Shapiro, S.~L. 2002, Astrophys. J. Lett., 572, L39,
  \dodoi{10.1086/341516}

\bibitem[{Shibata {et~al.}(2016)Shibata, Uchida, \&
  Sekiguchi}]{Shibata:2016vxy}
Shibata, M., Uchida, H., \& Sekiguchi, Y. 2016, Astrophys. J., 818, 157,
  \dodoi{10.3847/0004-637X/818/2/157}

\bibitem[{Siegel {et~al.}(2022)Siegel, Agarwal, Barnes, Metzger, Renzo, \&
  Villar}]{Siegel:2021ptt}
Siegel, D.~M., Agarwal, A., Barnes, J., {et~al.} 2022, Astrophys. J., 941, 100,
  \dodoi{10.3847/1538-4357/ac8d04}

\bibitem[{Stegmann {et~al.}(2025)Stegmann, Olejak, \&
  de~Mink}]{Stegmann:2025cja}
Stegmann, J., Olejak, A., \& de~Mink, S.~E. 2025.
\newblock \doarXiv{2507.15967}

\bibitem[{Takahashi {et~al.}(2018)Takahashi, Yoshida, \& Umeda}]{Takahashi2018}
Takahashi, K., Yoshida, T., \& Umeda, H. 2018, The Astrophysical Journal, 857,
  111, \dodoi{10.3847/1538-4357/aab95f}

\bibitem[{{Takahashi} {et~al.}(2016){Takahashi}, {Yoshida}, {Umeda},
  {Sumiyoshi}, \& {Yamada}}]{2016MNRAS.456.1320T}
{Takahashi}, K., {Yoshida}, T., {Umeda}, H., {Sumiyoshi}, K., \& {Yamada}, S.
  2016, \mnras, 456, 1320, \dodoi{10.1093/mnras/stv2649}

\bibitem[{Tanikawa {et~al.}(2025)Tanikawa, Liu, Wu, Fujii, \&
  Wang}]{Tanikawa:2025fxw}
Tanikawa, A., Liu, S., Wu, W., Fujii, M.~S., \& Wang, L. 2025.
\newblock \doarXiv{2508.01135}

\bibitem[{{Tassoul}(1978)}]{1978trs..book.....T}
{Tassoul}, J.-L. 1978, {Theory of rotating stars}

\bibitem[{{Timmes}(1999)}]{1999ApJS..124..241T}
{Timmes}, F.~X. 1999, \apjs, 124, 241, \dodoi{10.1086/313257}

\bibitem[{{Timmes} \& {Swesty}(2000)}]{2000ApJS..126..501T}
{Timmes}, F.~X., \& {Swesty}, F.~D. 2000, \apjs, 126, 501,
  \dodoi{10.1086/313304}

\bibitem[{Uchida {et~al.}(2019)Uchida, Shibata, Takahashi, \&
  Yoshida}]{Uchida:2018ago}
Uchida, H., Shibata, M., Takahashi, K., \& Yoshida, T. 2019, Astrophys. J.,
  870, 98, \dodoi{10.3847/1538-4357/aaf39e}

\bibitem[{{Zeldovich} \& {Novikov}(1971)}]{1971reas.book.....Z}
{Zeldovich}, Y.~B., \& {Novikov}, I.~D. 1971, {Relativistic astrophysics.
  Vol.1: Stars and relativity}

\end{thebibliography}

\end{document}